\documentclass{article}

\usepackage{arxiv}

\usepackage[utf8]{inputenc} % allow utf-8 input
\usepackage[T1]{fontenc}    % use 8-bit T1 fonts
\usepackage{hyperref}       % hyperlinks
\usepackage{url}            % simple URL typesetting
\usepackage{booktabs}       % professional-quality tables
\usepackage{amsfonts}       % blackboard math symbols
\usepackage{nicefrac}       % compact symbols for 1/2, etc.
\usepackage{microtype}      % microtypography
\usepackage{lipsum}

\usepackage{graphicx}
\usepackage{amsmath,amssymb}
\pdfoutput=1
\usepackage{tabularx}
\usepackage{changepage}

\usepackage{textcomp,marvosym}
\usepackage[right]{lineno}
\usepackage{microtype}
\DisableLigatures[f]{encoding = *, family = * }
\usepackage[table]{xcolor}
\usepackage{array}

\newcolumntype{+}{!{\vrule width 2pt}}

\newlength\savedwidth

\makeatletter

\title{Discrete $q$-exponential limit order cancellation time distribution}

\author{
  Vygintas~Gontis\thanks{Use footnote for providing further
    information about author (webpage, alternative
    address)---\emph{not} for acknowledging funding agencies.} \\
  Institute of Theoretical Physics and Astronomy\\
  Vilnius University\\
  Saul{\. e}tekio al. 3, 10257 Vilnius, Lithuania \\
  \texttt{vygintas@gontis.eu} \\
}

\begin{document}
\maketitle

\begin{abstract}
Modeling financial markets based on empirical data poses challenges in selecting the most appropriate models. Despite the abundance of empirical data available, researchers often face difficulties in identifying the best fitting model. Long-range memory and self-similarity estimators, commonly used for this purpose, can yield inconsistent parameter values, as they are tailored to specific time series models. In our previous work, we explored order disbalance time series from the broader perspective of fractional L'{e}vy stable motion, revealing a stable anti-correlation in the financial market order flow.
However, a more detailed analysis of empirical data indicates the need for a more specific order flow model that incorporates the power-law distribution of limit order cancellation times. When considering a series in event time, the limit order cancellation times follow a discrete probability mass function derived from the Tsallis {\textit{q}}%MDPI: italics or not. Please keep the format unified,  Gontis: Confirmed
-exponential distribution. The combination of power-law distributions for limit order volumes and cancellation times introduces a novel approach to modeling order disbalance in the financial markets. Moreover, this proposed model has the potential to serve as an example for modeling opinion dynamics in social systems. By tailoring the model to incorporate the unique statistical properties of financial market data, we can improve the accuracy of our predictions and gain deeper insights into the dynamics of these complex~systems.
\end{abstract}

% keywords can be removed

\section{Introduction}
In the realm of econometrics and econophysics, researchers have long explored the intriguing properties of long-range memory in natural and social systems, often characterized by self-similarity and power-law statistical distributions. In~financial markets, the~abundance of data related to volatility, trading activity, and~order flow has provided fertile ground for empirical investigations into long-range memory properties~\cite{Baillie1996JE,Engle2001QF,Plerou2001QF,Gabaix2003Nature,Ding2003Springer}. Various models with fractional noise have been proposed in econometrics to describe volatility time series~\cite{Ding1993JEmpFin,Baillie1996JE,Bollerslev1996Econometrics,Giraitis2009,Conrad2010,Arouri2012,Tayefi2012}. However, from~the perspective of econophysics, these models tend to lack sufficient microscopic reasoning and primarily serve as macroscopic descriptions of complex social systems, often based on assumptions of long-range memory. As~a result, predicting stock price movements, despite the application of advanced trading algorithms and machine learning techniques, remains an enduring challenge for researchers~\cite{Alec2015QF,Kumar2018IEEE,Zaznov2022Mathematics}.

To deepen our understanding of long-range memory in social systems, it becomes essential to compare the macroscopic modeling with empirical analyses. Our previous review~\cite{Kazakevicius2021Entropy} raised the crucial question of whether observed long-range memory in social systems is a result of genuine long-range memory processes or merely an outcome of the non-linearity of Markov processes. In~our endeavor to explore this question, we have reduced the macroscopic dynamics of financial markets to a set of stochastic differential equations (SDEs) and related them to a microscopic agent-based model capable of reproducing empirical probability density functions (PDFs) and power spectral densities (PSDs) of absolute returns~\cite{Kononovicius2013EPL,Gontis2014PlosOne,Gontis2017PhysA,Gontis2018PhysA}. Moreover, we have employed this model to interpret the scaling behavior of volatility return intervals~\cite{Gontis2016PhysA}. This approach could also find relevance in other social systems, where non-linear SDEs derived from agent-based models describing opinion or population dynamics lead to macroscopic descriptions~\cite{AbirDe2016dblp,Gontis2017Entropy}. Given these complexities, selecting the most appropriate model for interpreting empirical time series poses a significant~challenge.

A promising line of inquiry comes from the observation that market-order flows exhibit long-range persistence, attributed to the order-splitting behavior of individual traders~\cite{Lillo2005PhysRevE}. This finding reinforces the presence of genuine long-range memory in financial systems, as~recently confirmed in a comprehensive investigation~\cite{Kanazawa2023Arxiv}. Consequently, we anticipate the manifestation of persistence in the limit-order flow as well.

In our previous contribution, we explored order disbalance time series in financial markets from the perspective of fractional L\'{e}vy stable motion (FLSM) \cite{Gontis2022CNSNS}. Although~FLSM and auto-regressive fractionally integrated moving average (ARFIMA) processes offer generalized models for self-similar fractional time series~\cite{Burnecki2010PRE,Burnecki2014JStatMech,Burnecki2017ChaosSF}, they require more comprehensive approaches to explain the observed statistical properties of order disbalance in financial~markets.

In this study, we continue our analysis of limit order flow using LOBSTER data, see technical description~\cite{Huang2011Lobster} or short {description}
 \url{https://lobsterdata.com/info/HowDoesItWork.php}, aiming to demonstrate the possibility of simple, yet specific modeling of empirical time series. A~key innovation of our approach lies in the discovery of a general statistical distribution governing limit order cancellation times. We propose the application of Tsallis statistics, a~generalization of Boltzmann--Gibbs statistics~\cite{Tsallis2009BJP,Tsallis2017Entropy,Stosic2019PhysA}, to~fit the histograms of limit order cancellation times. Remarkably, the~distribution's parameters for various stocks and time periods appear close, suggesting a universal nature of the observed statistical~property.

To augment the identification of the limit order cancellation times' $q$-exponential distribution as a new stylized fact in limit order statistical properties, we further consider the assumption of limit order flow as fractional L\'{e}vy noise (FLN). Empirical data analysis supports the plausibility of such an assumption. Ultimately, we present a relatively straightforward model with empirical grounding, serving as an artificial model defining statistical properties of order flow and disbalance in financial markets. This new approach addresses certain contradictions uncovered in previous investigations from the perspective of FLSM~\cite{Gontis2022CNSNS}.

 The rest of this paper is organized as follows: Section~\ref{sec:disc-q-Exponential} introduces the discrete Tsallis $q$-exponential distribution; Section~\ref{sec:CancelTimes} defines and investigates the empirical statistical properties of limit order cancellation times; Section~\ref{sec:OrderFlow} examines a modified version of the order disbalance time series and its statistical properties; Section~\ref{sec:LO-series} studies the statistical properties of the limit order submission sequence; Section~\ref{sec:Model} introduces the artificial order disbalance model, and~finally, we conclude the results and summarize our findings.

\section{The Discrete Tsallis \boldmath{$q$}-Exponential~Distribution \label{sec:disc-q-Exponential}}

Let us introduce the discrete Tsallis $q$-exponential distribution to capture the probability mass function (PMF) for discrete variables $k=1,2,3,\dots$ based on the continuous Tsallis $q$-exponential probability density function (PDF) \cite{Tsallis2009BJP}, given by:
\begin{equation}
P_{\lambda,q}(x)=(2 - q) \lambda (1 - (1 - q) x \lambda)^{\frac{1}{(1 - q)}},
\label{Eq:q-Exponential}
\end{equation}
where $(1 - (1 - q) x \lambda)>0$. First, we explore an approach using the survival function of the $q$-exponential distribution~\cite{Bercher2008PhysA}:
\vspace{-3pt}
\vspace{-3pt}
\vspace{-3pt}  
\begin{equation}
SP_{\lambda,q}(x)=(1 +(q-1) x \lambda)^{\frac{2 - q}{1 - q}}.
\label{Eq:SP}
\end{equation}

\vspace{-3pt}
 
With
 this survival function, we define the PMF of the discrete $q$-exponential distribution as:
 \vspace{-3pt}
 \vspace{-3pt}
 \vspace{-3pt}   
\begin{equation}
P_{\lambda,q}^{(ds)}(k)=SP_{\lambda,q}(k-1)-SP_{\lambda,q}(k)=(1 + (q-1) (k-1) \lambda)^{\frac{2 - q}{1 - q}} - (1 + (q-1) k \lambda)^{\frac{2 - q}{1 - q}}.
\label{Eq:PMFfromSF}
\end{equation}

The PMF \eqref{Eq:PMFfromSF} is normalized for $k=1,2,\dots\infty$. Interestingly, this power-law PMF converges to the geometric distribution as $q\rightarrow 1$, which aligns with expectations that we have one more $q$-generalization of the geometric distribution~\cite{Matsuzoe2011WS,Yalcin2014JCAM}. We can express this convergence as:
\begin{equation}
\lim_{q \to 1} P_{\lambda,q}^{(ds)}(k) = \exp^{-y \lambda} (\exp^\lambda-1) = (1-p)^{k-1} p,
\label{Eq:q-limit}
\end{equation}
where we denote $p=1-\exp^{-\lambda}$. This result confirms the suitability of the PMF \eqref{Eq:PMFfromSF} as a Tsallis $q$-generalization of the geometric distribution, making it a potential candidate for modeling event waiting times in social and other complex systems.

The PMF for the discrete variable $k$ often is defined as follows,~\cite{Nekoukhou2012CSTM}:
\begin{equation}
P_{\lambda,q}^{(d)}(k)=\frac{P_{\lambda,q}(k)}{\sum_{i=1}^{i=\infty} P_{\lambda,q}(i)}.
\label{Eq:disc-PMF}
\end{equation}

After substitution of Equation~\eqref{Eq:q-Exponential} into Equation~\eqref{Eq:disc-PMF} we obtain a version of the discrete Tsallis $q$-exponential distribution as:

\vspace{-3pt}
\vspace{-3pt}
\vspace{-3pt}   
\begin{equation}
P_{\lambda,q}^{(d)}(k)=\frac{\{1+(1-q) \lambda k\}^{\frac{1}{1-q}}}{((q-1) \lambda)^{\frac{1}{1-q}}HurwitzZeta\left[\frac{1}{q-1},1+\frac{1}{\lambda (q-1)}\right]}.
\label{Eq:PMFwithHurZeta}
\end{equation}

The explicit form of the limit of this PMF when $q\rightarrow 1$ is currently unclear; thus, we will use Equation~\eqref{Eq:PMFfromSF} instead of Equation~\eqref{Eq:PMFwithHurZeta} in this investigation.

\section{Cancellation Times of Limit Orders in the Order Flow of Financial~Markets  \label{sec:CancelTimes}}
In this section, we analyze the cancellation times of limit orders in the order flow of financial markets using the LOBSTER data for all NASDAQ-traded stocks \cite{Huang2011Lobster}. The~LOB data that LOBSTER reconstructs originates from NASDAQ's Historical TotalView-ITCH{files} 
 (\url{http://nasdaqtrader.com}
). Here we construct the daily time series of order flow from 3 to 31 August 2020, a~total of 21 working days. This data exemplifies an empirical social system appropriate for investigating power-law statistical properties. The statistical properties of the cancellation times of limit orders are of particular interest in this study as they play a vital role in understanding order flow and disbalance dynamics.

We retrieve LOBSTER data files: message.csv and orderbook.csv for each selected trading day and ticker (stock). These files contain the complete list of events causing an update of LOB up to the ten levels of prices. Any event $j$ changing the LOB state has a time value $t_j=j$ counted in the event space. Thus, we deal with a discrete time scale and avoid daily seasonality related to the fluctuating activity of traders. Every limit order has its identification code; therefore, it is straightforward to pair
limit order submission and cancellation events. Let us define the limit order cancellation time as the time difference between the order cancellation and submission event. Seeking a simplified approach to the order flow modeling, we consider order cancellation and execution as equivalent events leading to the complete deletion of the previously submitted limit order. In~the LOBSTER message.csv file, these events are denoted as event types 3 and~4.

\begin{figure}[h]

\includegraphics[width=0.96\textwidth]{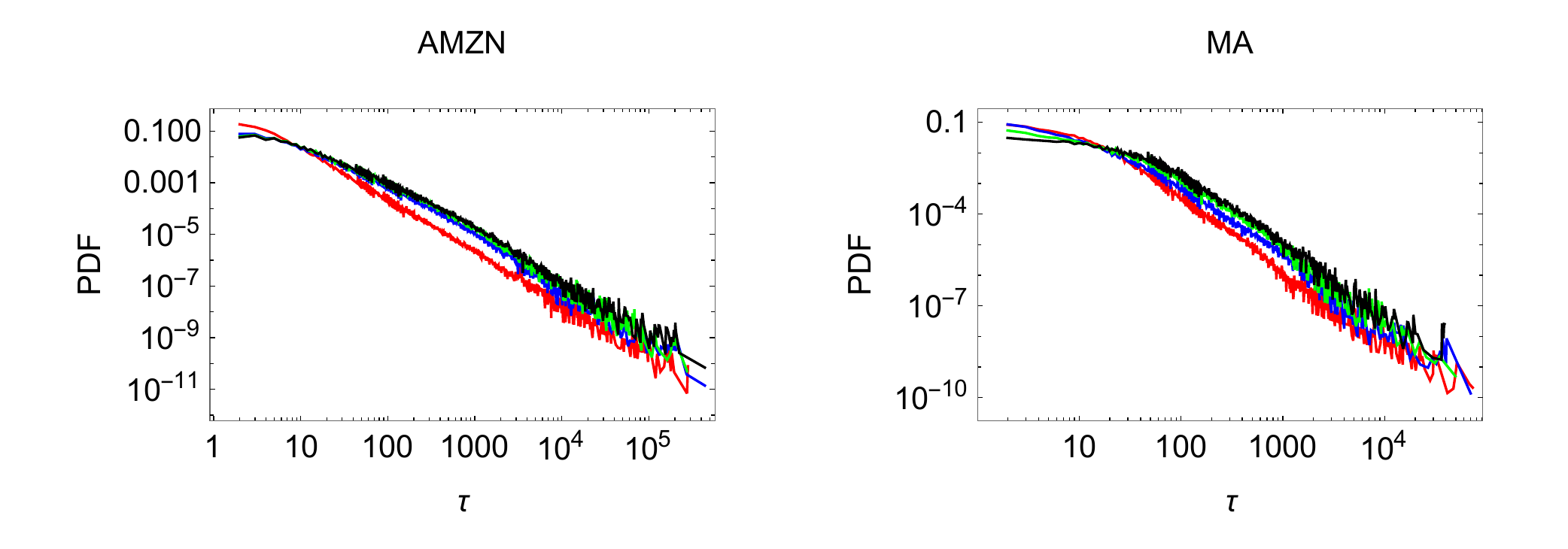}
\caption{Examples of the cancellation time histograms (PDFs) for the AMZN and MA stocks. Empirical PDFs for the four price levels are calculated for the joint period of 21 trading days. The~plots are red for the first level, blue for the second, green for the third, and~black for the~fourth. \label{fig1}}
\end{figure}

We can calculate the list of all cancellation times $\tau_i$ from empirical data or group this data according to the price levels or the volume of the limit orders. The~dependence of the cancellation time PDF on the level of price is weak. The~two sub-figures in Figure~\ref{fig1} illustrate the empirical PDFs of the cancellation times calculated separately for the four price levels of the two stocks AMZN and MA. It is worth noting that the dependence of the cancellation time PDF on the price level is weak, with~slight differences observed between the first price level (red) and the subsequent levels (blue, green, and~black).

\vspace{-3pt}
\vspace{-1pt}  

\begin{figure}[h]

\includegraphics[width=\textwidth]{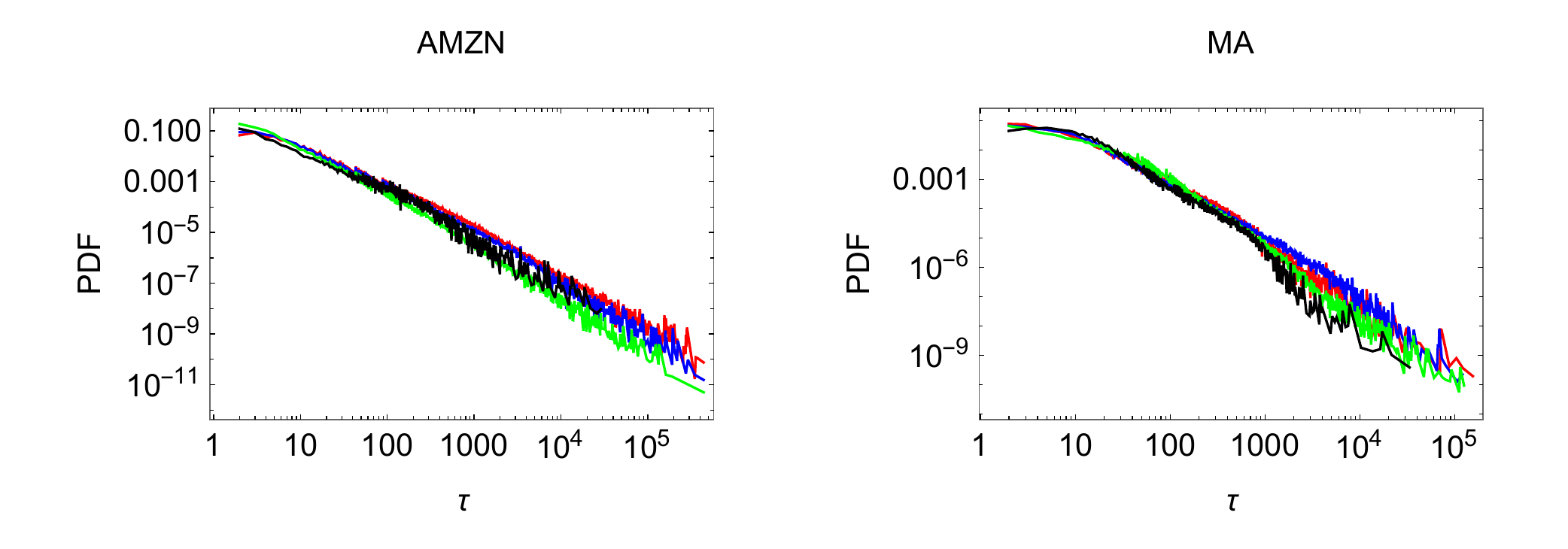}

\caption{Examples of cancellation time histograms (PDFs) for the AMZN and MA stocks. Empirical PDFs for the four intervals of limit order volumes are calculated for the joint period of 21 trading days. (red) $v_i\leq 3$; (blue) $3<v_i\leq 23$; (green) $23<v_i\leq 123$; (black) $123<v_i\leq 623$.\label{fig2}}
\end{figure}

We group the cancellation times $\tau_i$ into four categories based on the volumes $v_i$ of the limit orders: (1) $v_i\leq 3$; (2) $3<v_i\leq 23$; (3) $23<v_i\leq 123$; (4) $123<v_i\leq 623$. Each group provides sufficient $\tau_i$ values from the joint period of 21 trading days, enabling the calculation of histograms and evaluation of empirical probability density functions (PDFs). The two sub-figures in Figure~\ref{fig2} illustrate the empirical PDFs of AMZN and MA stock's cancellation times calculated separately for the four groups of limit order volumes. Once again, the~empirical distributions show weak dependence on the limit order volumes, reinforcing our modeling assumption of independence from price and volume.

\vspace{-7pt} 

Based on the independence assumption, we fit the empirical histograms of total limit order cancellation times submitted on the stock market for ten stocks (NVDA, HD, AMZN, NFLX, MA, LLY, TSLA, ADBE, V, JNJ) using the PMF \eqref{Eq:PMFfromSF}. The~fitting is performed using the Maximum Likelihood Estimator (MLE) method \cite{Shalizi2007MaximumLE}. Data for every stock is joined from 21~trading day of August 2020. Only limit orders up to the ten levels of prices are included in this MLE calculation. Table~\ref{table2} provides the calculated $\lambda$ and $q$ values for each stock from the joint empirical histograms.

\begin{figure}[h]

\includegraphics[width=0.75\textwidth]{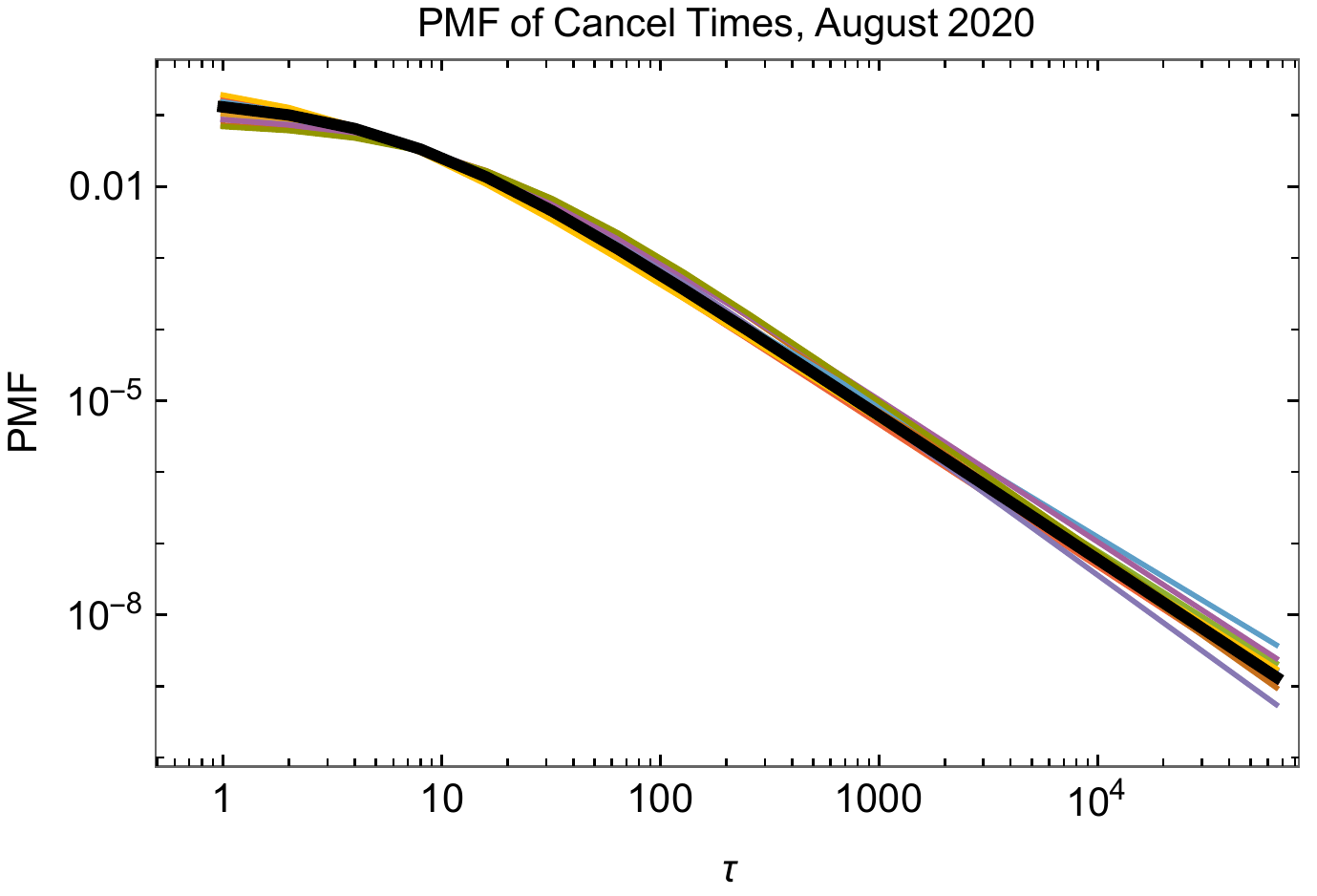}

\caption{Fitted PMFs \eqref{Eq:PMFfromSF} of the joined one-month cancellation times for the ten stocks NVDA, HD, AMZN, NFLX, MA, LLY, TSLA, ADBE, V, and JNJ. The thin color lines represent PMFs for the individual stocks, and~the thick black line represents the PMF with mean values of parameters $\lambda=0.3$ and $q=1.5$. Data for each MLE calculation is joined from 21 trading day of August~2020.\label{fig3}}
\end{figure}

The mean values of parameters for these ten stocks are $\lambda=0.3$ and $q=1.5$. Figure~\ref{fig3} demonstrates the close fitting PMFs for all ten stocks. The~thin color lines represent PMFs for individual stocks, while the thick black line represents the PMF with mean values \mbox{of parameters.}

\vspace{-3pt} 
 \vspace{-5pt} 
\vspace{-3pt} 
\begin{table}[h] 
\centering
\caption{
Parameters of PMF \eqref{Eq:PMFfromSF} $\{\lambda,q\}$ calculated using MLE for the cancellation times of ten stocks NVDA, HD, AMZN, NFLX, MA, LLY, TSLA, ADBE, V, and JNJ. Data for each MLE calculation is joined from 21 trading day of August 2020.}
\begin{tabularx}{\textwidth}{>{\centering\arraybackslash}X>{\centering\arraybackslash}X>{\centering\arraybackslash}X>{\centering\arraybackslash}X>{\centering\arraybackslash}X>{\centering\arraybackslash}X>{\centering\arraybackslash}X>{\centering\arraybackslash}X>{\centering\arraybackslash}X>{\centering\arraybackslash}X>{\centering\arraybackslash}X}
\toprule
\multicolumn{1}{X}{{ \textbf{Exp.}}} & \multicolumn{1}{X}{{ \textbf{NVDA}}} & \multicolumn{1}{X}{{ \textbf{HD}}} & \multicolumn{1}{X}{{\textbf{AMZN}}} & \multicolumn{1}{X}{{\textbf{NFLX}}} & \multicolumn{1}{X}{{\textbf{MA}}} & \multicolumn{1}{X}{{\textbf{LLY}}} & \multicolumn{1}{X}{{\textbf{TSLA}}} & \multicolumn{1}{X}{{ \textbf{ADBE}}} & \multicolumn{1}{X}{{~~~\textbf{V}}} & \multicolumn{1}{X}{{~~\textbf{JNJ}}}\\ \midrule
$\lambda$ & $0.38$ & $0.22$ & $0.41$ & $0.42$ & $0.17$ & $0.15$ & $0.4$ & $0.5$ & $0.19$ & $0.14$  \\  
$q$   & $1.51$ & $1.48$ & $1.52$ & $1.5$ & $1.45$ & $1.46$ & $1.54$ & $1.52$ & $1.5$ & $1.47$   \\ \bottomrule
\end{tabularx}
\label{table2}
\end{table}

\section{Live Limit Orders and Order~Disbalance  \label{sec:OrderFlow}}
The analysis of limit order cancellation times revealed intriguing empirical properties with remarkably close values of parameters for different stocks. This suggests the assumption that the observed probability mass function (PMF) might be stable over time, which can be valuable in constructing simplified order flow and disbalance models. To~achieve this objective, we reconstruct the order disbalance time series from the data in the message files. We adopt an alternative approach to form disbalance time series to explore more aspects of potential memory effects in empirical order disbalance time series.

As mentioned in Section~\ref{sec:CancelTimes}, the~LOBSTER data's message files contain sufficient information about limit orders submitted to the exchange, including their prices, volumes, event times of submission, cancellation, and~full execution. With~this information, we can construct a list of live limit orders at any discrete event step. We use the notation $LO\{i1, i2, v_{i1, i2}\}$ to denote a limit order, where $i1$ represents the order submission event time, $i2$ is the order cancellation or full execution time, and~$v_{i1, i2}$ is the volume (positive for buy limit orders and negative for sell limit orders). To~simplify the modeling approach, we ignore the indexing of price levels, as~it is not crucial for the investigation of the time series. We include limit orders up to the tenth level of prices on both the buy and sell sides. Using this notation, we can rewrite the order disbalance as follows:

\begin{equation}
X(j)=\sum_{i1\leqslant j < i2}v_{i1,i2}=\sum_{i=1}^{j} Y(i).
\label{eq:order-disbalance}
\end{equation}

Here, the~first sum is over all the live limit orders, including all the limit order volumes $v_{i1,i2}$ submitted before event $j$ and waiting for cancellation or execution. A~sequence of limit order submissions of length $N$ generates a series of order disbalance $X(j)$ of length $2N$ since each submission is paired with a cancellation or execution event.

This modified definition of order disbalance series provides a slightly simplified version compared to the previous approach used in~\cite{Gontis2022CNSNS}, which relied on the data from the orderbook.csv file. By~excluding order book events related to the partial execution of orders, we can focus on exploring the statistical properties of the time series with a simplified model. The~mean squared displacement (MSD) and Hurst exponents are evaluated for these empirical time series, considering two levels of random (reshuffled) time series. The~first level involves reshuffling only the empirical sequence of volumes $v_{i1,i2}$, yielding new series $X_{Rv}(j)$ and $Y_{Rv}(j)$. This reshuffling destroys the correlation contained in the limit order submission sequence. The~second level involves additional complete reshuffling of all increments $Y_R(j)=Random[Y(i)]$, leading to the memory-less time series $X_R(j)$, where the anti-correlation arising from the limit order cancellation events is destroyed as~well.

Table~\ref{table3} presents the MSD and Hurst exponents calculated for the order disbalance $X(j)$ of ten stocks NVDA, HD, AMZN, NFLX, MA, LLY, TSLA, ADBE, V, and~JNJ. The~Hurst exponents for different series are listed and evaluated using AVE and Higuchi's method. The~empirical analysis shows minor changes in scaling parameters compared to the results in~\cite{Gontis2022CNSNS}.

\begin{table}[h] 
\centering
\caption{
The MSD and Hurst's exponents calculated using the same equations as in~\cite{Gontis2022CNSNS} for the order disbalance $X(j)$, Equation~\eqref{eq:order-disbalance}, of~the ten stocks NVDA, HD, AMZN, NFLX, MA, LLY, TSLA, ADBE, V, and JNJ. Evaluated exponents are listed in the first column as follows: $\lambda_{MSD}$ is the exponent of sample MSD; $\lambda_{Rv}$ is the exponent for the series $X_{Rv}(j)$ with a randomized sequence of volumes; $H_{AV}$ is $H$ of series $X(j)$ evaluated using AVE; $H_{AVRv}$ is the same for the series $X_{Rv}(j)$ with a randomized sequence of volumes; $H_{AVR}$ is the same for the series $X_{R}(j)$ with a randomized sequence of $Y(j)$; $H_{Hig}$ is $H$ of series $X(j)$ evaluated using Higuchi's method; $H_{HigRv}$ is the same for the series $X_{Rv}(j)$ with a randomized sequence of volumes; $H_{HigR}$ is the same for the series $X_{R}(j)$ with a randomized sequence of $Y(j)$.}
\begin{tabularx}{\textwidth}{m{0.7cm}>{\centering\arraybackslash}X>{\centering\arraybackslash}X>{\centering\arraybackslash}X>{\centering\arraybackslash}X>{\centering\arraybackslash}X>{\centering\arraybackslash}X>{\centering\arraybackslash}X>{\centering\arraybackslash}X>{\centering\arraybackslash}X>{\centering\arraybackslash}X}
\midrule
\multicolumn{1}{X}{{\textbf{Exp.}}} & \multicolumn{1}{X}{{\textbf{NVDA}}} & \multicolumn{1}{X}{{~~\textbf{HD}}} & \multicolumn{1}{X}{{\textbf{AMZN}}} & \multicolumn{1}{X}{{\textbf{NFLX}}} & \multicolumn{1}{X}{{~\textbf{MA}}} & \multicolumn{1}{X}{{~\textbf{LLY}}} & \multicolumn{1}{X}{{\textbf{TSLA}}} & \multicolumn{1}{X}{{\textbf{ADBE}}} & \multicolumn{1}{X}{{~~\textbf{V}}} & \multicolumn{1}{X}{{~\textbf{JNJ}}}\\ \midrule
{$\lambda_{MSD}$} & $0.82$ & $1.01$ & $0.79$ & $0.86$ & $0.92$ & $1.05$ & $0.89$ & $0.88$ & $1.09$ & $1.10$  \\  
{\scriptsize $\lambda_{Rv}$} & $0.90$ & $0.93$ & $0.90$ & $0.88$ & $0.99$ & $0.94$ & $0.90$ & $0.86$ & $0.91$ & $0.94$   \\ 
{\scriptsize $H_{AV}$}    & $0.32$ & $0.21$ & $0.19$ & $0.21$ & $0.17$ & $0.2$ & $0.26$ & $0.26$ & $0.30$ & $0.31$   \\ 
{\scriptsize $H_{AVRv}$}   & $0.22$ & $0.16$ & $0.18$ & $0.18$ & $018$ & $0.18$ & $0.22$ & $0.18$ & $0.22$ & $0.21$   \\ 
{\scriptsize $H_{AVR}$}   & $0.54$ & $0.49$ & $0.54$ & $0.51$ & $0.48$ & $0.51$ & $0.50$ & $0.50$ & $0.50$ & $0.49$   \\ 
{\scriptsize $H_{Hig}$}   & $0.33$ & $0.22$ & $0.22$ & $0.24$ & $0.2$ & $0.22$ & $0.28$ & $0.28$ & $0.33$ & $0.32$  \\ 
{\scriptsize $H_{HigRv}$}  & $0.22$ & $0.18$ & $0.20$ & $0.18$ & $0.18$ & $0.18$ & $0.22$ & $0.29$ & $0.23$ & $0.21$   \\ 
{\scriptsize $H_{HigR}$}  & $0.54$ & $0.49$ & $0.50$ & $0.51$ & $0.49$ & $0.50$ & $0.51$ & $0.50$ & $0.53$ & $0.49$   \\ \bottomrule
\end{tabularx}
\label{table3}
\end{table}  
  
Our previous exploration of the order flow from the perspective of FLSM or ARFIMA models~\cite{Gontis2022CNSNS} has sparked new questions and raised concerns about the applicability of these models. Notably, while considering the time series $X(j)$ as an accumulated ARFIMA(0,d,0) process provided valuable insights, it was observed that empirical order disbalance time series exhibit strict boundedness. This discrepancy indicates the existence of diffusion reversion mechanisms that fall outside the scope of the ARFIMA model.

\begin{figure}[h]
\includegraphics[width=0.85\textwidth]{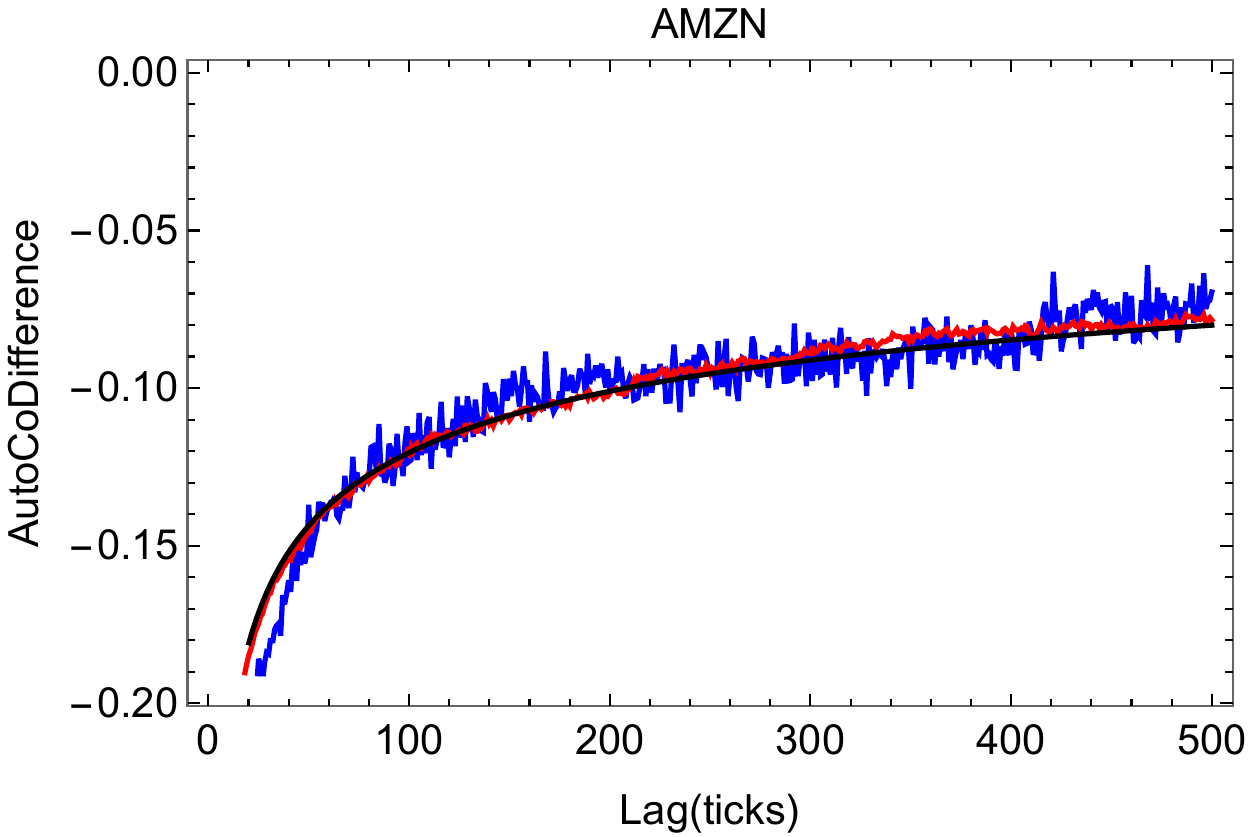}
\caption{Auto-{codifference}, \eqref{eq:CD-sample}, 
 from a single empirical trajectory $Y(j)$ of AMZN stock traded on 3 August 2020, blue line. Black line, the~fit of empirical auto-codifference by $-0.388 t^{-0.254}$. Red line, auto-codifference calculated from a single trajectory of fractional L\'{e}vy noise with parameters $d=0.375$ and $\alpha=2.0$
\label{fig4}}
\end{figure}

Another critical challenge that emerged during our analysis is related to the auto-codifference of empirical $Y(j)$ time series~\cite{Wylomanska2015PhysA}. The~order disbalance increments $Y(j)$ demonstrate clear auto-dependence, as~shown in Figure~\ref{fig4}. However, fitting the empirical auto-codifference with the expected theoretical form $\sim D t^{\eta}$ derived for fractional L'{e}vy noise~\cite{Astrauskas1991LMR} presented difficulties. Qualitatively, the~asymptotic auto-codifference of the ARFIMA(0,d,0) process only matched the empirical data for memory parameter values in the region of $d \simeq 0.375$. However, within~this region, the~accumulated ARFIMA process exhibited strong unbounded behavior, which is inconsistent with the empirical order disbalance time series.

To gain a more specific empirical perspective, we investigated the potential application of the FLSM approach to the order disbalance series, assuming that the distribution of limit order volumes for the most liquid stocks followed a stable L'{e}vy distribution with parameter $2>\alpha>1$ \cite{Gontis2022CNSNS}. However, this assumption turned out to be very approximate due to the distinctive resonance structure in the PDF of volumes. This observation led us to reevaluate the limit order flow from a different angle.

One specific issue arising from the FLSM perspective was related to the sample mean squared displacement (MSD), which has an exponent $\lambda_{MSD}=2 d+1$ for large samples and lags~\cite{Burnecki2010PRE}. However, these values of the parameter $d$ defined by the empirical $\lambda_{MSD}$ in Table~\ref{table3} range from $d=-0.11$ for AMZN to $d=0.07$ for JNJ. These values contradict the empirical analysis in~\cite{Gontis2022CNSNS}, which suggested $d=H_{AV}-H_{AVR} \simeq -0.3$ for all stocks.
 
\begin{figure}[h]
\includegraphics[width=0.9\textwidth]{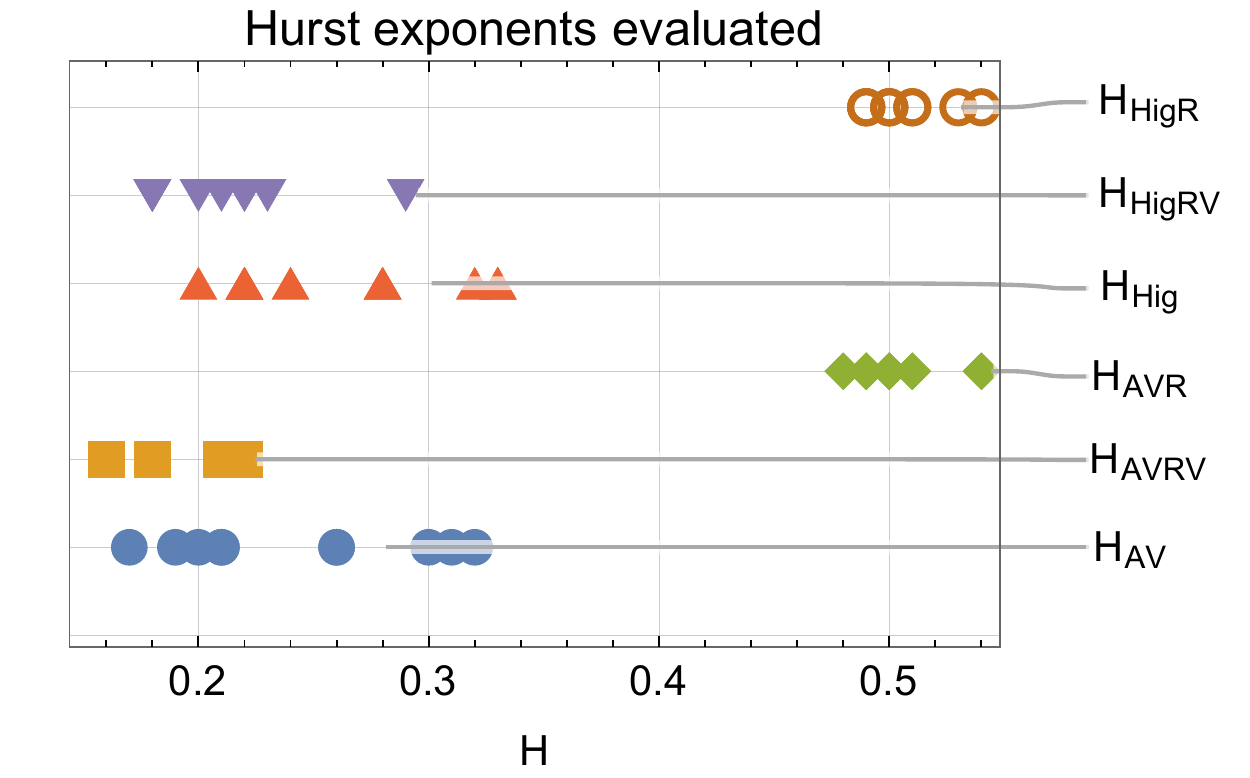}
\caption{Comparison of exponents calculated for the empirical and randomized time series. All rows have $10$ values corresponding to the stocks investigated in this contribution. Scaling parameters are labeled as defined in the text and Table~\ref{table3}. 
\label{fig5}}
\end{figure}

To address limitations of the previous approach and gain deeper insights into the auto-dependence of increments $Y(j)$, we introduced a modified order disbalance definition, Equation~\eqref{eq:order-disbalance}. The~pairing of limit order increments with opposite signs due to cancellation times assumed to follow a $q$-exponential PMF provides a new quantitative model contributing to the observed auto-dependence. This new mechanism of auto-dependence is fundamentally different from the fractionally integrated increments in the ARFIMA{0,d,0} series.
Thus, the~simplification introduced here is helpful for the more precise interpretation of memory effects in the order disbalance time series. In~Figure~\ref{fig5}, we visualize the results of empirical analysis presented in Table~\ref{table3}.

The auto-dependence in the $Y(j)$ series can also have other origins; for example, some dependence can arise from the original sequence of the limit order volumes $v_{i1,i2}$. Such memory effects in market order flow have been investigated in~\cite{Lillo2005PhysRevE,Toth2015JEDC,Zaznov2022Mathematics}, and~competing interpretations have been provided. To~evaluate the possible contribution of the auto-dependence in the limit order volume sequence, we use the random reshuffling procedure of the limit order volumes to obtain $X_{Rv}(j)$ and $Y_{Rv}(j)$ series with zero volume correlation. The~evaluated Hurst exponents of the $X_{Rv}(j)$ series, see Table~\ref{table3} and Figure~\ref{fig5}, are slightly shifted to the side of smaller values. Notably, time series $X_{Rv}(j)$ remained strictly bounded, in~contrast to the unbounded behavior observed in $X_R(j)$, suggesting that the diffusion of the order disbalance series is self-reverted due to every limit order being canceled or executed. The~observed auto-dependence in $Y(j)$ and its origin from both the cancellation times and the sequence of limit order volumes raise intriguing questions about the underlying mechanisms of market dynamics. Further theoretical consideration and empirical studies may help uncover the nature of the long-range memory in order disbalance and contribute to a more comprehensive understanding of market order flow.

\section{Time Series of Limit Order~Submissions  \label{sec:LO-series}}
It is essential to investigate the statistical properties of time series that comprise only a sequence of limit order submissions to the market, denoted as $X_L(j)$,
\begin{equation}
X_L(j)=\sum_{i=1}^{j} v_i = \sum_{i=1}^{j} Y_L(i), 
\label{eq:limit-order-sequence}
\end{equation}
where $v_i$ represents the volume of the submitted limit order, and~order cancellation or execution is not included in the series. In~this case, we can confidently consider the series from the perspective of FLSM, as~the order flow remains uninterrupted by cancellation events. The~analysis results are provided in Table~\ref{table4}.

\begin{table}[h] 
\centering
\caption{Evaluated memory parameter $d$ for the limit order flow time series $X_L(j)$, as~defined in Equation~\eqref{eq:limit-order-sequence}, for~ten stocks: NVDA, HD, AMZN, NFLX, MA, LLY, TSLA, ADBE, V, and~JNJ. Three different methods are used for evaluation, listed in the first column: $d_{MSD}$ is evaluated from the exponent of sample MSD, $\lambda_{MSD}=2 d_{MSD}+1$; $d_{CD}$ is evaluated from the sample auto-codifference, as~defined in Equation~\eqref{eq:CD-sample}; $d_H$ is evaluated as $H_{AVE}-H_{AVER}$ from the sample series $X_L(j)$ and $X_{LR}(j)$ using AVE. The~last line presents the empirical values of the stability parameter $\alpha$ obtained by fitting the tails of volume histograms with the L'{e}vy stable~distribution.}
\begin{tabularx}{\textwidth}{>{\centering\arraybackslash}X>{\centering\arraybackslash}X>{\centering\arraybackslash}X>{\centering\arraybackslash}X>{\centering\arraybackslash}X>{\centering\arraybackslash}X>{\centering\arraybackslash}X>{\centering\arraybackslash}X>{\centering\arraybackslash}X>{\centering\arraybackslash}X>{\centering\arraybackslash}X}
\midrule
\multicolumn{1}{X}{\textbf{Exp.}} & \multicolumn{1}{X}{\textbf{NVDA}} & \multicolumn{1}{X}{~~\textbf{HD}} & \multicolumn{1}{X}{\textbf{AMZN}} & \multicolumn{1}{X}{\textbf{NFLX}} & \multicolumn{1}{X}{~\textbf{MA}} & \multicolumn{1}{X}{~\textbf{LLY}} & \multicolumn{1}{X}{\textbf{TSLA}} & \multicolumn{1}{X}{\textbf{ADBE}} & \multicolumn{1}{X}{~~\textbf{V}} & \multicolumn{1}{X}{~~\textbf{JNJ}}\\ \midrule
\textbf{$d_{MSD}$} & $0.30$ & $0.18$ & $0.32$ & $0.29$ & $0.18$ & $0.14$ & $0.28$ & $0.28$ & $0.20$ & $0.20$   \\ 
{\scriptsize $d_{CD}$} & $0.29$ & $0.23$ & $0.30$ & $0.29$ & $0.19$ & $0.20$ & $0.34$ & $0.30$ & $0.26$ & $0.25$   \\  
{\scriptsize $d_H$} & $0.32$ & $0.21$ & $0.38$ & $0.32$ & $0.21$ & $0.15$ & $0.35$ & $0.34$ & $0.20$ & $0.19$   \\  
{\scriptsize $\alpha$}    & $1.80$ & $1.80$ & $1.79$ & $1.80$ & $1.80$ & $1.80$ & $1.80$ & $1.79$ & $1.79$ & $1.79$    \\ \bottomrule
\end{tabularx}
\label{table4}
\end{table}

The first method used to evaluate the parameter $d$ is based on the assumption that the series $X_L(j)$ exhibits FLSM-like behavior, with~the memory parameter $d_{MSD}$ derived from the sample mean squared displacement (MSD) exponent $\lambda_{MSD}=2 d_{MSD}+1$, as~referenced in \cite{Burnecki2010PRE,Gontis2022CNSNS}. The second method, $d_{CD}$, involves calculating the sample auto-codifference $CD(t)$ with lag $t$ of series $Y_L(j)$, as~defined in Equation~\eqref{eq:CD-sample}. It is worth noting that this method is sensitive to the evaluation of parameter $\alpha$ due to its reliance on the asymptotic form of auto-codifference for fractional L'{e}vy noise, as~given in Equation~\eqref{eq:CD-asymptotic}, leading to the definition $d_{CD}=\gamma/\alpha+1-1/\alpha$ \cite{Wylomanska2015PhysA}.
\begin{equation}
CD(t)=\ln \frac{\sum_{j=1+t}^{j=N} i Y_L(j)+\sum_{j=1}^{j=N-t}-i Y_L(j)}{N \sum_{j=1+t}^{j=N} i (Y_L(j)-Y_L(j-t))}, 
\label{eq:CD-sample}
\end{equation}
where $i$ denotes imaginary units and $N$ is the length of the series. Note that this method is sensitive to the evaluation of parameter $\alpha$, as~we use the asymptotic form
\begin{equation}
CD(t) \sim t^{\gamma}=t^{\alpha H-\alpha},
\label{eq:CD-asymptotic}
\end{equation}
of auto-codifference for the fractional L\'{e}vy noise, see~\cite{Wylomanska2015PhysA}.
The third method, $d_H$, relies on the relation used in~\cite{Gontis2022CNSNS}, where $d_H=H_{AV}-H_{AVR}$, assuming that the time series $X_L(j)$ is fractional L'{e}vy stable motion-like.

Evaluation of memory parameter $d$ using three different methods and results in the Table~\ref{table4} support the idea that the limit order series $X_L(j)$ is FLSM-like. Fluctuations of memory parameters between methods are considerably smaller than fluctuations between different stocks. The~major problem in this consideration remains the assumption of the limit order volume distribution according to the L\'{e}vy stable distribution. We fit the $\alpha$ parameter of this distribution to only the tail part $v_i>35$ of the empirical histogram. Results provided in the Table~\ref{table4} show very stable values for all stocks $\alpha \simeq 1.8$.

\section{Artificial Order Disbalance Time~Series \label{sec:Model}}
We propose an artificial order disbalance time series model that captures the main observations of this study. The~model involves two random sequences:
(a) A sequence of limit order volumes generated as ARFIMA\{0,d,0\}\{$\alpha,N,v_{\text{max}}$\}, where $d$ is the memory parameter, $\alpha$ is the stability index, $N$ is the length of the sequence, and~$v_{\text{max}}$ defines the maximum possible absolute value of the volume, selected from the observed empirical data.
(b) A corresponding sequence of limit order cancellation times with the same length $N$ generated using the probability mass function (PMF) $P_{\lambda,q}^{(ds)}(k)$ defined by Equation~\eqref{Eq:PMFfromSF}.

With these two independent sequences, we can calculate the model sequence of events $X_M(j)=\sum_{i=1}^{i=j}Y_M(i)$ defined by $v_{i1,i2}^{mod}$, which includes cancellation and execution events. This generated random sequence represents the artificial analog of order disbalance time series introduced in Equation~\eqref{eq:order-disbalance}, and it is used for the empirical analysis. We choose the artificial model parameters aiming to reproduce the empirical data: $\alpha=1.8$; $N=200000$; $v_{max}=1000$; $\lambda=0.42$; $q=1.5$.

We compare results generated by the artificial series $Y_M(i)$ and $X_M(j)$ with those from the empirical series. In~Table~\ref{table5}, we present the MSD and Hurst exponents, together with the memory parameter $d_{CD}$, for~the artificially generated series and empirical series of stocks MA, NFLX, and~TSLA.

The results in the Table~\ref{table5} are averaged over five realizations of artificial series and five daily empirical series. The~model results with $d=0.2$ closely match the series of stock MA, while the model results with $d=0.3$ are similar to the empirical series of NFLX and TSLA. Thus, we conclude that the empirically established $q$-exponential nature of the limit order cancellation times helps to reconstruct the properties of the order disbalance time series $X(j)$. Such reconstructions are crucial for interpreting order disbalance time series from the perspective of~FLSM. 

\begin{table}[h] 
\small
\caption{
Sample series evaluated MSD and Hurst exponents, along with memory parameter $d_{CD}$, for the artificially generated series and the empirical series of stocks MA, NFLX, and TSLA. $\lambda(X)$ represents the sample MSD exponent evaluated for the series $X(j)$; $\lambda(X_L)$ is the same for the series of limit volumes; $H(X)$ is the Hurst exponent evaluated using AVE for the series $X(j)$; $H(X_R)$ is the same for the reshuffled series $X_R(J)$; $H(X_L)$ is the same for the series of limit volumes $X_L(j)$; $H(X_{LR})$ is the same for the reshuffled limit volume series $X_{LR}(j)$; and $d_{CD}$ is the memory parameter evaluated from %Please check intended meaning has been retained
the sample auto-codifference of series $Y(i)$.}
\setlength{\tabcolsep}{3.07mm}
{\begin{tabular}{cccccccc}
\toprule
\multicolumn{1}{c}{\textbf{Series}} & \multicolumn{1}{c}{\boldmath{$\lambda(X)$}} & \multicolumn{1}{c}{\boldmath{$\lambda(X_v)$}} & \multicolumn{1}{c}{\boldmath{$H(X)$}} & \multicolumn{1}{c}{\boldmath{$H(X_R)$}} & \multicolumn{1}{c}{\boldmath{$H(X_v)$}} & \multicolumn{1}{c}{\boldmath{$H(X_{LR})$}} & \multicolumn{1}{c}{\boldmath{$d_{CD}$}} \\ \midrule
{\scriptsize Model} $d=0.2$ & $0.97$ & $1.36$ & $0.20$ & $0.55$ & $0.72$ & $0.54$ & $0.18$    \\  
{\scriptsize MA}            & $0.92$ & $1.37$ & $0.17$ & $0.48$ & $0.71$ & $0.50$ & $0.19$    \\  
{\scriptsize Model} $d=0.3$ & $1.00$ & $1.60$ & $0.24$ & $0.54$ & $0.83$ & $0.55$ & $0.22$    \\  
{\scriptsize NFLX}          & $0.86$ & $1.58$ & $0.21$ & $0.51$ & $0.83$ & $0.51$ & $0.29$    \\ 
{\scriptsize TSLA}          & $0.89$ & $1.56$ & $0.26$ & $0.50$ & $0.87$ & $0.52$ & $0.34$    \\ \bottomrule
\end{tabular}}
\label{table5}
\end{table}

\begin{figure}[h]

%\begin{adjustwidth}{-\extralength}{0cm}
\centering 
\includegraphics[width=0.9\textwidth]{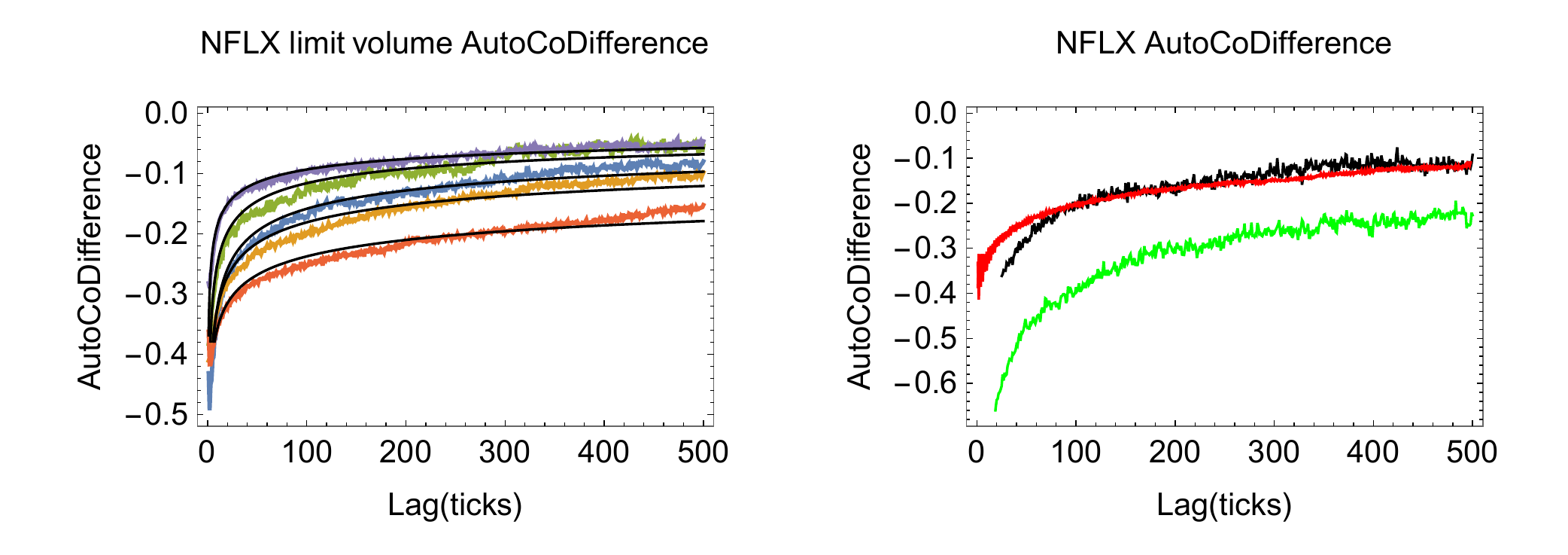}
%\end{adjustwidth} 
\caption{Sample auto-codifference of the NFLX time series, Equation~\eqref{eq:CD-sample}. In~the left sub-figure auto-codifference of the empirical time series $Y_L(i)$, the collar curves represent five daily series, and~th eblack lines are the best fit by $C \tau^\gamma$. In~the right sub-figure: red represents the auto-codifference averaged over five daily empirical NFLX time series $Y(i)$; black represents the auto-codifference averaged over five realizations of model series $Y(i)$; and green represents the auto-codifference averaged over \mbox{five realizations} of model limit volume series $Y_L(i)$.} 
\label{fig6}
\end{figure}

In Figure~\ref{fig6}, we demonstrate the application of sample auto-codifference~\cite{Wylomanska2015PhysA}, as~defined in Equation~\eqref{eq:CD-sample}, for~the empirical and artificial model time series. The~left sub-figure shows the NFLX auto-codifference of the limit order flow $Y_L(i)$ for the first five trading days of August 2020, along with the best fit by the asymptotic curve from Equation~\eqref{eq:CD-asymptotic}. The~fluctuations in defined parameters $\gamma=\{0.30,0.25,0.33,0.18,0.30\}$ and \mbox{$d_{CD}=\{0.28, 0.31, 0.26, 0.34, 0.28\}$} are considerable for the daily series, but~the average $d$ is close to $0.3$. The~right sub-figure compares three auto-codifference curves, averaged over five sample series related to NFLX data. The~red curve represents the auto-codifference of the empirical order disbalance increment series $Y(j)$, the~black curve of the corresponding order disbalance artificial model series with $d=0.3$, and~the green curve of the artificial limit order flow series $Y_L(i)$. The~similar behavior of all three curves and good correspondence between the empirical and synthetic series indicate the usefulness of auto-codifference in the research of persistence in financial and other social~systems.

An intriguing result is that the auto-codifference of two very different processes in terms of self-similarity exhibits the expected behavior. Positively correlated fractional L'{e}vy noise-like series $Y_L(i)$ display a similar auto-codifference as series $Y(i)$ that exhibit anti-persistence from a self-similarity perspective ($H<<0.5$). For~example, our preliminary investigation of empirical series based only on the limit order signs shows that the persistence of limit order flow disappears when cancellation and execution of orders are included in the~series.

In conclusion, the~artificial order disbalance time series model provides valuable insights into the persistence and memory properties of the series. The~comparison with empirical data demonstrates the usefulness of the model and supports the conclusion that the $q$-exponential nature of limit order cancellation times contributes to the observed persistence in order disbalance time series. The~application of auto-codifference further enhances our understanding of the self-similarity behavior in these financial systems, opening up new avenues for research in this domain.

\section{Discussion and~Conclusions  \label{sec:Conclusions}}
In this study, we have delved into the statistical properties of limit order cancellation times in financial markets to better understand the peculiarities of order disbalance time series from the perspective of fractional L'{e}vy stable motion (FLSM). Our previous investigation of order disbalance time series within the framework of FLSM yielded contradictory conclusions~\cite{Gontis2022CNSNS}. However, empirical time series often exhibit specific characteristics that necessitate careful consideration during empirical analysis. To~address the question of why order disbalance time series in financial markets are strictly bounded, we have focused on the statistical properties of limit order cancellation times, treating them as discrete~events.

\textls[+25]{To this end, we have introduced the concept of a discrete $q$-exponential distribution, presented in Equation~\eqref{Eq:PMFfromSF}, as~a $q$-extension of the geometric distribution, based on the theoretical foundations of generalized Tsallis statistics~\cite{Tsallis1988-ku}. This distribution allows for a better fit of empirical limit order cancellation times, revealing their weak sensitivity to order sizes and price levels. Remarkably, the~parameters of the fitted discrete $q$-exponential PMF, $\lambda=0.3$, and~$q=1.5$, have proven consistent across ten stocks and trading days analyzed. Building on this unique statistical property of cancellation times, we model and empirically investigate limit order flow and order disbalance time~series.}

The clear distinction between the series of limit order flow $X_L(j)$ and the series of order disbalance $X(j)$, which includes order cancellations, is essential in this research. Limit order flow series display persistence and remain unbounded, making them FLSM-like. On~the other hand, order disbalance series, which includes order cancellations and executions, is bounded and exhibits anti-persistence. It is important to acknowledge that limit order flow in financial markets serves as a prime example of time series requiring thorough empirical analysis to validate the use of econometric methods for time series~analysis.

By combining fractional L'{e}vy stable limit order flow with the $q$-exponential cancellation time distribution, we propose a relatively straightforward model of order disbalance in financial markets. This model also serves as an illustrative example of the broader approach to modeling opinion dynamics in various social systems. Our research highlights the significance of social system modeling to ensure the proper utilization of formal mathematical~methods.

In summary, our study contributes to a better understanding of order disbalance time series and their memory effects in financial markets. The~incorporation of the discrete $q$-exponential distribution for modeling cancellation times provides valuable insights into the persistence of the order disbalance time series and helps address the question of their boundedness. Furthermore, the~combination of FLSM and the $q$-exponential distribution proves to be a promising approach for modeling social systems, which can be explored further in future~research.

In conclusion, the~statistical properties of limit order cancellation times and their impact on order disbalance time series have shed light on the dynamics of financial markets. This study not only enhances our understanding of complex financial systems but also highlights the importance of empirical analysis when applying mathematical methods to social system modeling. By~bridging the gap between theory and empirical observations, we contribute to the development of more accurate models and deeper insights into the behavior of financial markets and social systems as a whole.

\section{Abbreviations}
The following abbreviations are used in this manuscript:\\
 
\noindent 
\begin{tabular}{@{}ll}
ARFIMA & Auto--regressive fractionally integrated moving average\\
AVE & Absolute Value estimator\\
FBM & Fractional Brownian motion\\
FGN & Fractional Gaussian noise\\
FLSM & fractional L\`{e}vy stable motion\\
MSD & Mean squared displacement\\
PDF & Probability density function\\
PMF & Probability mass function
\end{tabular}

%\begin{adjustwidth}{-\extralength}{0cm} 


\begin{thebibliography}{999}

\bibitem{Baillie1996JE}
{B}aillie, R.T.; Bollerslev, T.; Mikkelsen, H.O.
\newblock Fractionally integrated generalized autoregressive conditional
heteroskedasticity.
\newblock {\em \mbox{J. Econom.}} {\bf 1996}, {\em 74},~3--30. [\href{http://doi.org/10.1016/S0304-4076(95)01749-6}{CrossRef}]
%MDPI: Please DO NOT change/revert the form of references in Reference Section, they have been completed layout and ready for publication. Please just provide the detailed information if required in the comments below. Or please provide the website links and accessed date (Day Month Year) if you cannot provide detailed information. Gontis: Confirmed


\bibitem{Engle2001QF}
Engle, R.; Patton, A.
\newblock What good is a volatility model?
\newblock {\em Quant. Financ.} {\bf 2001}, {\em 1},~237--245. [\href{http://dx.doi.org/10.1088/1469-7688/1/2/305}{CrossRef}]

\bibitem{Plerou2001QF}
Plerou, V.; Gopikrishnan, P.; Gabaix, X.; Amaral, L.; Stanley, H.
\newblock Price fluctuations, market activity and trading volume.
\newblock {\em \mbox{Quant. Financ.}} {\bf 2001}, {\em 1},~262--269. [\href{http://dx.doi.org/10.1088/1469-7688/1/2/308}{CrossRef}]

\bibitem{Gabaix2003Nature}
Gabaix, X.; Gopikrishnan, P.; Plerou, V.; Stanley, H.E.
\newblock A theory of power law distributions in financial market fluctuations.
\newblock {\em Nature} {\bf 2003}, {\em 423},~267--270. [\href{http://dx.doi.org/10.1038/nature01624}{CrossRef}] [\href{http://www.ncbi.nlm.nih.gov/pubmed/12748636}{PubMed}]

\bibitem{Ding2003Springer}
Rangarajan, G.; Ding, M.; (Eds.) {\em Processes with Long-Range Correlations: Theory and Applications,} {\em Lecture Notes
in Physics};
Springer:  {Berlin/Heidelberg, Germany,}
2003; Volume 621, p. XVIII, 398.

\bibitem{Ding1993JEmpFin}
Ding, Z.; Granger, C.W.J.; Engle, R.F.
\newblock A long memory property of stock market returns and a new model.
\newblock {\em J. Empir. Financ.} {\bf 1993}, {\em 1},~83--106. [\href{http://dx.doi.org/10.1016/0927-5398(93)90006-D}{CrossRef}]

\bibitem{Bollerslev1996Econometrics}
Bollerslev, T.H.-O.;~Mikkelsen, H.O.
\newblock Modeling and pricing long-memory in stock market volatility.
\newblock {\em J. Econom.} {\bf 1996}, {\em 73},~151--184.

\bibitem{Giraitis2009}
Giraitis, L.; Leipus, R.; Surgailis, D.
\newblock {ARCH}($\infty$) models and long memory. In {\em Handbook of
Financial Time Series}; Anderson, T.G., Davis, R.A., Kreis, J., Mikosh, T.,
Eds.; Springer: Berlin/Heidelberg, Germany, 2009; pp. 71--84.
\newblock.\_3. [\href{http://dx.doi.org/10.1007/978-3-540-71297-8_3}{CrossRef}]

\bibitem{Conrad2010}
Conrad, C.
\newblock Non-negativity conditions for the hyperbolic GARCH model.
\newblock {\em J. Econom.} {\bf 2010}, {\em 157},~441--457. [\href{http://dx.doi.org/10.1016/j.jeconom.2010.03.045}{CrossRef}]

\bibitem{Arouri2012}
Arouri, M.E.H.; Hammoudeh, S.; Lahiani, A.; Nguyen, D.K.
\newblock Long memory and structural breaks in modeling the return and
volatility dynamics of precious metals.
\newblock {\em  Q. Rev. Econ. Financ.} {\bf 2012}, {\em
52},~207--218. [\href{http://dx.doi.org/10.1016/j.qref.2012.04.004}{CrossRef}]

\bibitem{Tayefi2012}
Tayefi, M.; Ramanathan, T.V.
\newblock An overview of {FIGARCH} and related time series models.
\newblock {\em Austrian J. Stat.} {\bf 2012}, {\em 41},~175--196. [\href{http://dx.doi.org/10.17713/ajs.v41i3.172}{CrossRef}]

\bibitem{Alec2015QF}
Kercheval, A.N.; Zhang, Y.
\newblock Modelling high-frequency limit order book dynamics with support
vector machines.
\newblock {\em Quant. Financ.} {\bf 2015}, {\em 15},~1315--1329. [\href{http://dx.doi.org/10.1080/14697688.2015.1032546}{CrossRef}]

\bibitem{Kumar2018IEEE}
Kumar, I.; Dogra, K.; Utreja, C.; Yadav, P.
\newblock A Comparative Study of Supervised Machine Learning Algorithms for
Stock Market Trend Prediction.
\newblock In Proceedings of the 2018 Second International Conference on
Inventive Communication and Computational Technologies ({ICICCT}), Coimbatore, India, 20--21 April 2018. [\href{http://dx.doi.org/10.1109/icicct.2018.8473214}{CrossRef}]

\bibitem{Zaznov2022Mathematics}
Zaznov, I.; Kunkel, J.; Dufour, A.; Badii, A.
\newblock Predicting Stock Price Changes Based on the Limit Order Book: A
Survey.
\newblock {\em Mathematics} {\bf 2022}, {\em 10}, 1234. [\href{http://dx.doi.org/10.3390/math10081234}{CrossRef}]

\bibitem{Kazakevicius2021Entropy}
Kazakevicius, R.; Kononovicius, A.; Kaulakys, B.; Gontis, V.
\newblock Understanding the Nature of the Long-Range Memory Phenomenon in
Socioeconomic Systems.
\newblock {\em Entropy} {\bf 2021}, {\em 23}, 1125. [\href{http://dx.doi.org/10.3390/e23091125}{CrossRef}]

\bibitem{Kononovicius2013EPL}
Kononovicius, A.; Gontis, V.
\newblock Three state herding model of the financial markets.
\newblock {\em EPL} {\bf 2013}, {\em 101},~28001. [\href{http://dx.doi.org/10.1209/0295-5075/101/28001}{CrossRef}]

\bibitem{Gontis2014PlosOne}
Gontis, V.; Kononovicius, A.
\newblock Consentaneous agent-based and stochastic model of the financial
markets.
\newblock {\em PLoS ONE} {\bf 2014}, {\em 9},~e102201. [\href{http://dx.doi.org/10.1371/journal.pone.0102201}{CrossRef}] [\href{http://www.ncbi.nlm.nih.gov/pubmed/25029364}{PubMed}]

\bibitem{Gontis2017PhysA}
Gontis, V.; Kononovicius, A.
\newblock Burst and inter-burst duration statistics as empirical test of
long-range memory in the financial markets.
\newblock {\em Phys. A} {\bf 2017}, {\em 483},~266--272. [\href{http://dx.doi.org/10.1016/j.physa.2017.04.163}{CrossRef}]

\bibitem{Gontis2018PhysA}
Gontis, V.; Kononovicius, A.
\newblock The consentaneous model of the financial markets exhibiting spurious
nature of long-range memory.
\newblock {\em Phys. {A}} {\bf 2018}, {\em 505},~1075--1083. [\href{http://dx.doi.org/10.1016/j.physa.2018.04.053}{CrossRef}]

\bibitem{Gontis2016PhysA}
Gontis, V.; Havlin, S.; Kononovicius, A.; Podobnik, B.; Stanley, H.E.
\newblock Stochastic model of financial markets reproducing scaling and memory
in volatility return intervals.
\newblock {\em Phys. A} {\bf 2016}, {\em 462},~1091--1102. [\href{http://dx.doi.org/10.1016/j.physa.2016.06.143}{CrossRef}]

\bibitem{AbirDe2016dblp}
De, A.; Valera, I.; Ganguly, N.; Bhattacharya, S.; Gomez{-}Rodriguez, M.
\newblock Learning and Forecasting Opinion Dynamics in Social Networks.
\newblock In Proceedings of the Advances in Neural Information Processing
Systems 29: Annual Conference on Neural Information Processing Systems 2016,  Barcelona, Spain,  5--10 December 2016; Lee, D.D., Sugiyama, M., von Luxburg,
U., Guyon, I., Garnett, R., Eds.; pp. 397--405.

\bibitem{Gontis2017Entropy}
Gontis, V.; Kononovicius, A.
\newblock Spurious memory in non-equilibrium stochastic models of imitative
behavior.
\newblock {\em Entropy} {\bf 2017}, {\em 19},~387. [\href{http://dx.doi.org/10.3390/e19080387}{CrossRef}]

\bibitem{Lillo2005PhysRevE}
Lillo, F.; Mike, S.; Farmer, J.D.
\newblock Theory for long memory in supply and demand.
\newblock {\em Phys. Rev. E} {\bf 2005}, {\em 71},~066122. [\href{http://dx.doi.org/10.1103/PhysRevE.71.066122}{CrossRef}]

\bibitem{Kanazawa2023Arxiv}
Sato, Y.; Kanazawa, K.
\newblock Exact solution to a generalised Lillo-Mike-Farmer model with
heterogeneous order-splitting strategies. \emph{arXiv} \textbf{2023}, arXiv:2306.13378.
\url{https://doi.org/10.48550/arXiv.2306.13378}.

\bibitem{Gontis2022CNSNS}
Gontis, V.
\newblock Order flow in the financial markets from the perspective of the
Fractional Lévy stable motion.
\newblock {\em Commun. Nonlinear Sci. Numer. Simul.}
{\bf 2022}, {\em 105},~106087.
\newblock.: 10.1016/j.cnsns.2021.106087. [\href{http://dx.doi.org/10.1016/j.cnsns.2021.106087}{CrossRef}]

\bibitem{Burnecki2010PRE}
Burnecki, K.; Weron, A.
\newblock Fractional {L}evy stable motion can model subdiffusive dynamics.
\newblock {\em Phys. Rev. E} {\bf 2010}, {\em 82},~021130. [\href{http://dx.doi.org/10.1103/PhysRevE.82.021130}{CrossRef}]

\bibitem{Burnecki2014JStatMech}
Burnecki, K.; Weron, A.
\newblock Algorithms for testing of fractional dynamics: {A} practical guide to
{ARFIMA} modelling.
\newblock {\em J. Stat. Mech.} {\bf 2014}, {\em
2014},~P10036. [\href{http://dx.doi.org/10.1088/1742-5468/2014/10/p10036}{CrossRef}]

\bibitem{Burnecki2017ChaosSF}
Burnecki, K.; Sikora, G.
\newblock Identification and validation of stable {ARFIMA} processes with
application to {UMTS} data.
\newblock {\em Chaos Solitons Fractals} {\bf 2017}, {\em 102},~456--466. [\href{http://dx.doi.org/10.1016/j.chaos.2017.03.059}{CrossRef}]

\bibitem{Huang2011Lobster}
Huang, R.; Polak, T.  {LOBSTER}: The limit Order Book Reconstructor.  \emph{Discussion Paper School of Business and Economics};
\newblock Technical Report; Humboldt Universitat zu Berlin:  Berlin, Germany, {2011}.
\newblock
\bibitem{Tsallis2009BJP}
Tsallis, C.
\newblock Nonadditive entropy and nonextensive statistical mechanics---An
overview after 20 years.
\newblock {\em Braz. J. Phys.} {\bf 2009}, {\em 39},~337--356. [\href{http://dx.doi.org/10.1590/S0103-97332009000400002}{CrossRef}]

\bibitem{Tsallis2017Entropy}
Tsallis, C.
\newblock Economics and finance: Q-{S}tatistical stylized features galore.
\newblock {\em Entropy} {\bf 2017}, {\em 19},~457. [\href{http://dx.doi.org/10.3390/e19090457}{CrossRef}]

\bibitem{Stosic2019PhysA}
Stosic, D.; Stosic, D.; Stosic, T.
\newblock Nonextensive triplets in stock market indices.
\newblock {\em Phys. A Stat. Mech. Appl.} {\bf
2019}, {\em 525},~192--198.
\newblock.: 10.1016/j.physa.2019.03.093. [\href{http://dx.doi.org/10.1016/j.physa.2019.03.093}{CrossRef}]

\bibitem{Bercher2008PhysA}
Bercher, J.F.; Vignat, C.
\newblock A new look at q-exponential distributions via excess statistics.
\newblock {\em Phys. A Stat. Mech.  Appl.} {\bf
2008}, {\em 387},~5422--5432. [\href{http://dx.doi.org/10.1016/j.physa.2008.05.038}{CrossRef}]

\bibitem{Matsuzoe2011WS}
Matsuzoe, H.; Ohara, A. Geometry for q-Exponential Families.
\newblock In {\em Recent Progress in Differential Geometry and Its Related
Fields}; World Scientific: Singapore, {2011}; pp. 55--71. [\href{http://dx.doi.org/10.1142/9789814355476_0004}{CrossRef}]

\bibitem{Yalcin2014JCAM}
Yalcin, F.; Eryilmaz, S.
\newblock \emph{q}-geometric and q-binomial distributions of order k.
\newblock {\em J. Comput. Appl. Math.} {\bf 2014},
{\em 271},~31--38. [\href{http://dx.doi.org/10.1016/j.cam.2014.03.025}{CrossRef}]

\bibitem{Nekoukhou2012CSTM}
Nekoukhou, V.; Alamatsaz, M.H.; Bidram, H.
\newblock A Discrete Analog of the Generalized Exponential Distribution.
\newblock {\em Commun. Stat. Theory Methods} {\bf 2012},
{\em 41},~2000--2013. [\href{http://dx.doi.org/10.1080/03610926.2011.555044}{CrossRef}]

\bibitem{Shalizi2007MaximumLE}
Shalizi, C.R.
\newblock Maximum Likelihood Estimation for \emph{q}-Exponential (Tsallis)
Distributions. \emph{arXiv} {\bf 2007}, arXiv:math/0701854.


\bibitem{Wylomanska2015PhysA}
Wyłomańska, A.; Chechkin, A.; Gajda, J.; Sokolov, I.M.
\newblock Codifference as a practical tool to measure interdependence.
\newblock {\em Phys. A Stat. Mech.  Appl.} {\bf
2015}, {\em 421},~412–429. [\href{http://dx.doi.org/10.1016/j.physa.2014.11.049}{CrossRef}]

\bibitem{Astrauskas1991LMR}
Astrauskas, A.; L\'{e}vy, J.B.; Taqqu, M.S.
\newblock The asymptotic dependence structure of the linear fractional L\'{e}vy
motion.
\newblock {\em Liet. Mat. Rink.} {\bf 1991}, {\em 31},~3--28.

\bibitem{Toth2015JEDC}
T{\'o}th, B.; Palit, I.; Lillo, F.; Farmer, J.D.
\newblock Why is equity order flow so persistent?
\newblock {\em J. Econ. Dyn. Control} {\bf 2015}, {\em 51},~218--239.

\bibitem{Tsallis1988-ku}
Tsallis, C.
\newblock Possible generalization of {Boltzmann-Gibbs} statistics.
\newblock {\em J. Stat. Phys.} {\bf 1988}, {\em 52},~479--487. [\href{http://dx.doi.org/10.1007/BF01016429}{CrossRef}]

\end{thebibliography}
\end{document}